\documentclass[aps,prb,preprint,showpacs,groupedaddress]{revtex4}

\usepackage{color} 
\usepackage{epsfig}
\preprint{BaCuGeO-2014}

\begin{document}
\DeclareGraphicsExtensions{.eps,.jpg,.png}
\input epsf
\title{Infrared phonon spectrum of the tetragonal helimagnet Ba$_2$CuGe$_2$O$_7$} 
\author {A. Nucara$^{1}$, W. S. Mohamed$^{2}$, L. Baldassarre$^{3}$, S. Koval$^{4}$, J. Lorenzana$^{5}$, R. Fittipaldi$^{6}$, G. Balakhrisnan$^{7}$, A. Vecchione$^{6}$, and P. Calvani$^{1}$}
\affiliation{$^{1}$CNR-SPIN and Dipartimento di Fisica,  Universit\`{a} di Roma ''La Sapienza'', P.le A. Moro 2, 00185 Roma, Italy}\
\affiliation{$^{2}$Dipartimento di Fisica,  Universit\`{a} di Roma ''La Sapienza'', P.le A. Moro 2, 00185 Roma, Italy}\
\affiliation{$^{3}$Center for Life NanoScience@Sapienza, Istituto Italiano di Tecnologia, Viale Regina Elena 291, 00185 Roma, Italy}\
\affiliation{$^{4}$Instituto de F\'{\i}sica Rosario, Universidad Nacional de Rosario, 27 de Febrero 210 Bis, 2000 Rosario, Argentina}\
\affiliation{$^{5}$CNR-ISC and Dipartimento di Fisica,  Universit\`a di Roma ''La Sapienza'', P.le A. Moro 2, 00185 Roma, Italy}\
\affiliation{$^{6}$CNR-SPIN and Dipartimento di Fisica "E. R. Caianiello", Via Giovanni Paolo II 132, 84084 Fisciano, Salerno, Italy}\
\affiliation{$^{7}$Department of Physics, University of Warwick, Coventry CV4 7AL, UK}\

\date{\today}

\begin{abstract}
The lattice dynamics of  Ba$_2$CuGe$_2$O$_7$, a compound which develops Dzyaloshinsky-Moriya (DM) helical magnetism below $T_N$ = 3.2 K, has been studied by measuring the infrared reflectivity of a single crystal  with the radiation polarized both in the $ab$ plane and along the $c$ axis of its tetragonal cell, from 7 K to 300 K. In this compound, where the unit cell has no inversion symmetry, fourteen $E$ phonon modes of the $ab$ plane, out of the eighteen predicted, and all the ten $B_2$ modes of the $c$ axis, have been observed. They have been assigned to the atomic motions by a comparison with shell-model calculations, which provided vibrational frequencies in good agreement with the experiment, while most calculated intensities turned to be much lower than the experimental values. This discrepancy has been tentatively explained by assuming strong electron-phonon interactions, a hypothesis supported by the failure of the $f$- sum rule if restricted to the phonon region. Indeed, we observe a remarkable increase in the oscillator strengths at $T$'s low but higher than $T_N$, which suggests that the dielectric constant of  Ba$_2$CuGe$_2$O$_7$ may increase at those temperatures.
\end{abstract}
\maketitle

\section{Introduction}

Thanks to the growing research effort on the magnetoelectric compounds - which below a critical temperature simultaneously display ferroelectricity and some kind of magnetic order - several mechanisms that can couple magnetism to a macroscopic polarization have been identified. Among them, one of the most interesting is the  Dzyaloshinsky-Moriya (DM) effect \cite{D,M}, which gives rise to helical spin structures through an exchange Hamiltonian which depends on the vector product between adjacent spins. Such magnetism can also be associated with an electric polarization $\vec P$ whose expression contains the same vectorial term \cite{Murakawa12}.

\begin{figure}[b]
\begin{center}
{\hbox{\epsfig{figure=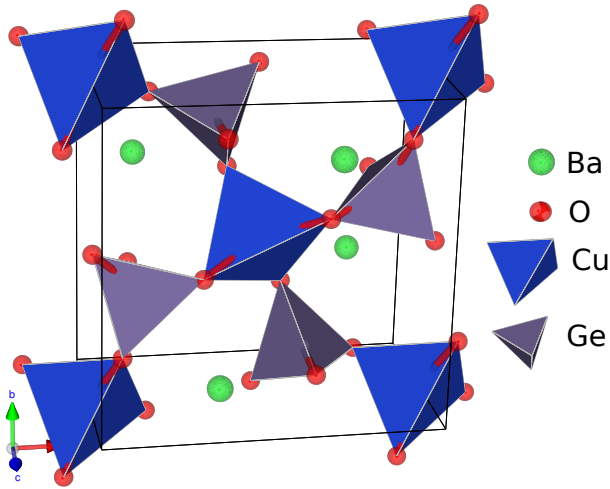,width=8cm}}}
\caption{Color online. Lattice structure of Ba$_2$CuGe$_2$O$_7$ (reelaborated from Ref. \onlinecite{Tovar98} with the graphical tools reported in Ref. \onlinecite{Momma}). The oxygen tetrahedra contain a copper atom if blue, a germanium atom if grey.}
\label{structure}
\end{center}
\end{figure}

The compound here studied, Ba$_2$CuGe$_2$O$_7$ (BCGO), is the only member of Ba$_2$XGe$_2$O$_7$ family which develops helical magnetism at liquid helium temperatures via the DM mechanism \cite{Zheludev1,Zheludev2,Chovan}. Indeed, for $X$=Mn (spin $S$ = 5/2) and Co  ($S$ = 3/2) the insulator is antiferromagnetic (AF) below  N\'eel temperatures $T_N$ = 4.0 and 6.7 K, respectively \cite{Masuda,Zheludev03,Murakawa12}. Below $T_N$ these systems develop a magnetoelectricity due to the spin-dependent hybridization between the $d$ orbitals of the transition metal and the $p$ orbitals of oxygen \cite{Murakawa10,Picozzi}. Ba$_2$CuGe$_2$O$_7$ instead, having the Cu$^{2+}$ ion $S$ = 1/2,  is magnetically  much less anisotropic. Below $T_N$ = 3.2 K, it thus displays a quasi-AF cycloidal, incommensurate magnetism. However, despite the absence of a center of inversion symmetry in the crystal structure, BCGO does not develop  spontaneous ferroelectricity \cite{Zheludev03}.  Nevertheless, a macroscopic electric polarization can be induced in it by an external magnetic field \cite{Murakawa09}.

As the magnetoelectric properties of  Ba$_2$CuGe$_2$O$_7$ have been extensively discussed in Refs. \onlinecite{Murakawa09,Murakawa10,Murakawa12}, the present work is focused on its lattice dynamics, both from an experimental and from a theoretical point of view. We have thus measured  the reflectivity $R(\omega)$ of a single crystal of this compound,  from 80 to 6000 cm$^{-1}$ and from 7 to 300 K, with the radiation polarized both along the $a$ (or $b$) axis  and along the $c$ axis of the tetragonal unit cell.  For both polarizations, the real part $\sigma_1(\omega)$ of the optical conductivity extracted  from $R(\omega)$  displays in the far infrared a number of well defined phonon peaks, as Ba$_2$CuGe$_2$O$_7$ is an excellent insulator. The measured  frequencies and intensities have been compared with  the predictions of shell-model calculations, which started from the structural determinations by X-ray diffraction. BCGO was found \cite{Tovar98} to crystallize in the non-centrosymmetric tetragonal space group P\=42$_1$m, with lattice parameters $a = b$ = 0.8466 nm and $c$ = 0.5445 nm These values are presumably measured at room temperature, and no structural changes induced by temperature are reported in the literature at best of our knowledge. A schematic view of its unit cell - corresponding to two formula units - is reported in Fig. \ref{structure}, which reproduces that reported in Ref. \onlinecite{Zheludev2}. The  Ba$^{2+}$ planes, orthogonal to the $c$ axis, separate the layers made of corner-sharing GeO$_4$ and CuO$_4$ tetrahedra.  It is in the resulting two-dimensional square lattice of Cu$^{2+}$ ions, that the Cu spins interact through the DM mechanism and the helical magnetic structure takes place.  

\section{Experiment and results}

Single crystals of Ba$_2$CuGe$_2$O$_7$ were grown by the floating zone technique at a pressure of 3 bar in oxygen gas atmosphere with the growth speed of 0.5 mm/hour \cite{Fittipaldi}. The morphology, phase composition and purity of the grown crystals were inspected by high resolution x-ray diffraction and scanning electron microscopy combined with energy dispersive spectroscopy. The crystal orientation was determined by x-ray back-reflection Laue method showing well defined, 
neither distorted nor smeared-out spots. Moreover, the crystals used in this work were cut in such a way that the largest surface contained the $a$ (or $ b$) and the $c$- axes. This surface was finely polished with polycrystalline diamond suspension down to 0.3 $\mu$m in grain size while the orientation of the $a-c$ (or $b-c$) axes with respect to the sample edges  was assessed by electron back-scattered Kikuchi diffraction technique. The dimensions of the crystals employed in this work were approximately 3$\times$2$\times$1 mm. In the visible range they look transparent and colourless like a piece of glass.

The reflectivity $R(\omega)$ of this surface was measured with a rapid-scanning Michelson interferometer at a resolution of 2 cm$^{-1}$. Two different polarizations of the radiation field were used, one along the $c$ axis  and the other one perpendicular to the $c$ axis (the latter being  indicated, in the following, as belonging to the $ab$ plane) . The sample was mounted on the cold finger of  a helium-flow cryostat.  Even if the phonon region is limited to  850 cm$^{-1}$, $R(\omega)$ was measured  by a rapid-scanning interferometer up to 6000 cm$^{-1}$ in order to perform on it accurate Kramers-Kronig (KK) transformations. The reference was a gold film evaporated in-situ onto the sample. As the crystal is nearly transparent in the midinfrared, to avoid interference fringes its back surface was left unpolished and wedged \cite{App_Opt98} with respect to the front surface. The real part $\sigma_1(\omega)$ of the optical conductivity was then extracted from $R(\omega)$ by standard KK transforms. The reflectivity was extrapolated to $\omega \to \infty$  by an inverse power law of $\omega$, and to $\omega = 0$ by Lorentzian fits.

\begin{figure}[b]
\begin{center}
{\hbox{\epsfig{figure=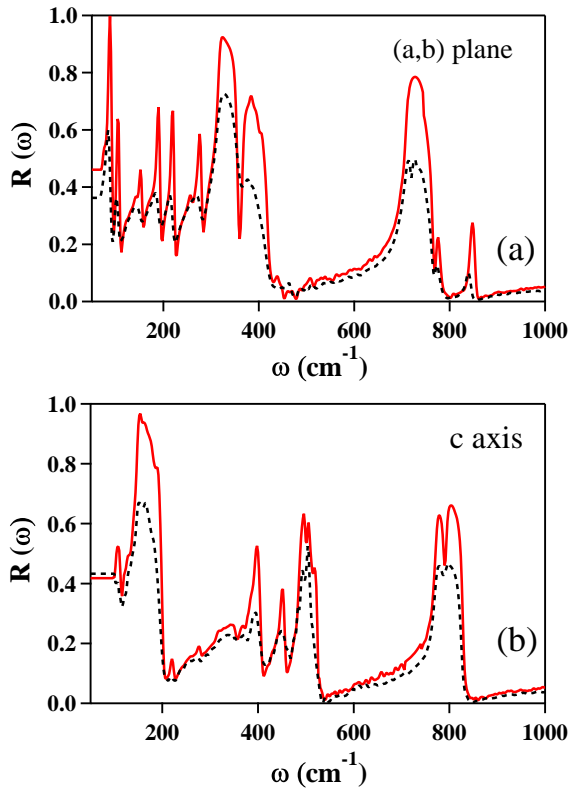,width=8cm}}}
\caption{Color online. Reflectivity of Ba$_2$CuGe$_2$O$_7$, in the far-infrared range of frequencies at 7 K (solid line) and 300 K (dashed line), with the radiation polarized in the $ab$ plane (a) and along the $c$ axis (b).}
\label{Rfir}
\end{center}
\end{figure}

The raw reflectivity data of Ba$_2$CuGe$_2$O$_7$ are shown in Fig. \ref{Rfir} in the frequency range of interest here, for  the radiation polarized  both in the $ab$ plane (a) and along the $c$ axis (b), either at 300 K and 7 K. No interesting effects are shown by the spectra at intermediate temperatures. Both spectra look very similar to the corresponding reflectivity spectra of Ba$_2$CoGe$_2$O$_7$, as  reported in the same frequency range in Ref. \onlinecite{Hutanu}. They are typical of an insulating crystal, with a comb of phonon lines in the far infrared, and a flat and very low reflectivity  (6-7 \%) in the midinfrared range (not shown). 

\begin{figure}[b]
\begin{center}
{\hbox{\epsfig{figure=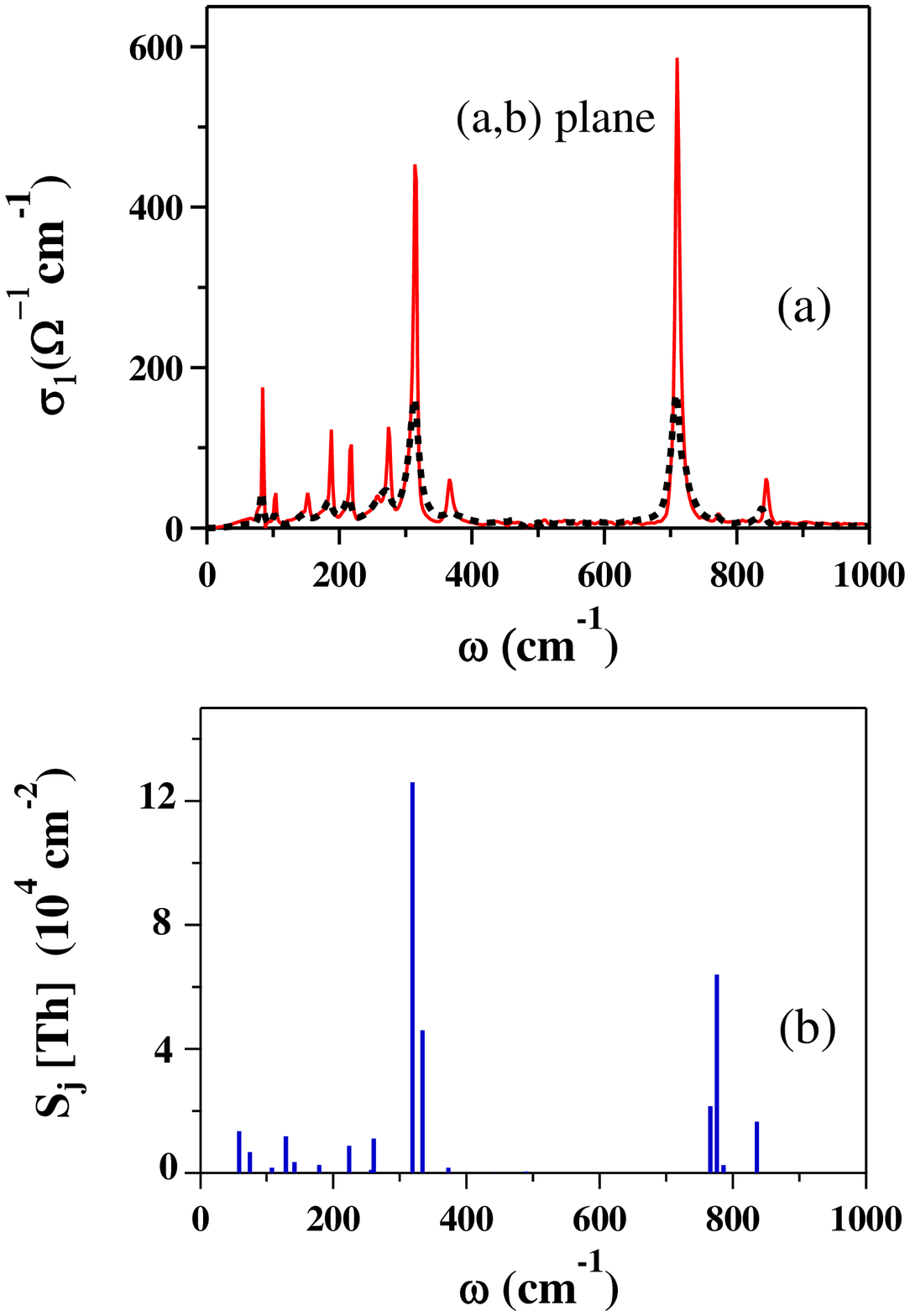,width=8cm}}}
\caption{Color online.  a) Optical conductivity of Ba$_2$CuGe$_2$O$_7$, in the far infrared range of frequencies at 7 K (solid line) and 300 K (dashed line), with the radiation polarized in the $ab$ plane. b) Shell-model results for the vibrational modes of the $ab$ plane. The bar length is the calculated strength $S_j$[th] in cm$^{-2}$.}
\label{sigma_ab}
\end{center}
\end{figure}

The optical conductivity extracted from the data of Fig. \ref{Rfir} is shown in Fig. \ref{sigma_ab} 
in the far-infrared region, both at the lowest and highest measured temperatures. We have detected 
fourteen transverse optical (TO) phonon modes with the radiation polarized in the $ab$ plane, out of the eighteen $E$ modes
predicted by a factor-group analysis of the P\=42$_1$m unit cell, and all the ten $B_2$ modes predicted along the $c$ axis. The absence of line frequencies common to both polarizations confirms that the polarizer was correctly oriented and that the sample was a single crystal.

\begin{figure}[b]
\begin{center}
{\hbox{\epsfig{figure=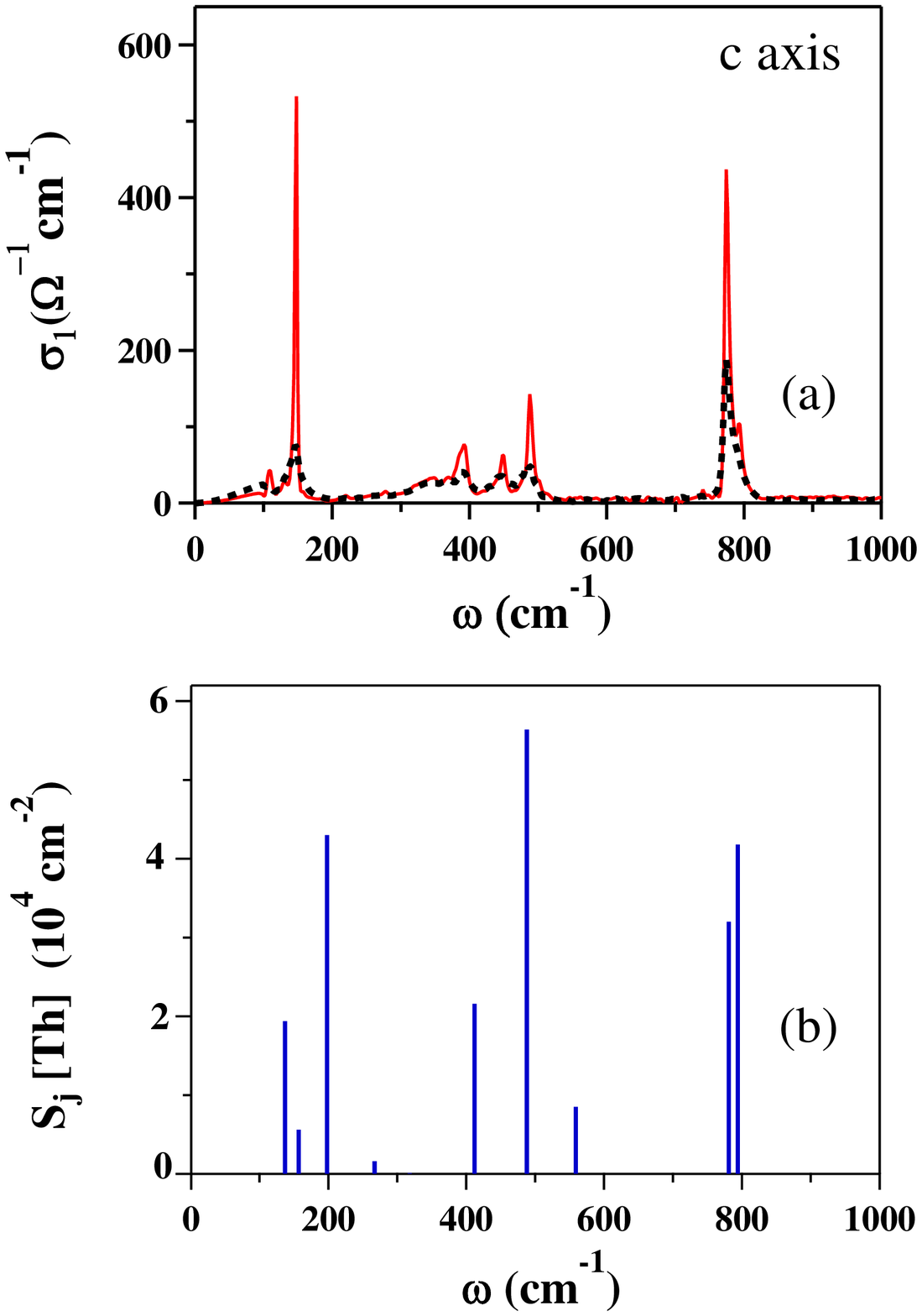,width=8cm}}}
\caption{Color online.  Optical conductivity of Ba$_2$CuGe$_2$O$_7$, in the far infrared range of frequencies at 7 K (solid line) and 300 K (dashed line), with the radiation polarized along the $c$ axis. 
b) Shell-model results for the vibrational modes along the $c$ axis. The bar length is the calculated strength $S_j$[th] in cm$^{-2}$.}
\label{sigma_c}
\end{center}
\end{figure}

\section{Comparison with theory}

In order to understand the complex phonon spectrum of BCGO, we have used  a shell model (SM) which enabled us to
perform lattice dynamical calculations and to compare the results with the measured $\sigma_1 (\omega)$.
The SM has been successfully applied to different compounds (oxides and hydrides) 
where the effects of the anion polarizability turned out to be important for a proper
description of the vibrational properties.\cite{koval92,koval96,lasave05,lasave09} 
The ionic polarizability is taken into account by considering electronic shells with 
charge $Y$ coupled harmonically by a force constant $k$ to an atomic core. 
The SM includes long-range Coulomb interactions between all 
charged shells and cores and shell-shell short-range interactions
arising from the wavefunction overlap between neighboring ions.
We have considered short-range interactions of the Born-Mayer type,
$A ~ exp(- \frac{r}{\rho} )$, for the Cu-O, Ba-O and Ge-O bonds.
The lattice constants and the atomic positions were taken from the experimental data of
Ref. \onlinecite{Tovar98}.

To calculate the oscillator frequencies and strengths we have fixed the total charges $Z$ of the ions to their nominal values, {\textit i.e.}, $Z_O$ = -2, $Z_{Cu}$ = 2, $Z_{Ba}$ = 2, and $Z_{Ge}$ = 4. The model contains 14 adjustable parameters. 
The initial parameter values were taken from  Refs. \onlinecite{braden02,tinte99} 
and were subsequently modified in order to fit the measured infrared phonon frequencies at the 
Brillouin zone center. The calculations were carried out with the help of the GULP code.\cite{gale97} 
Due to the complexity of the structure and consequently of the developed SM,  
the use of automatic searches for parameters included in the code turned out to be ineffective.\cite{lasave09} 
Thus, the adjustment of the parameters was made by hand. 
The final set of SM parameters which gave the best fit to data is shown in Table I.
The theoretical frequencies $\Omega_{j}$[th] of the transverse optical (TO) infrared phonons 
obtained with the model are shown in Table II for the $ab$ plane and in Table III 
for the $c$ axis. Also shown are the calculated oscillator strengths
$S_{j}$[th] for each $\j$-th TO infrared mode.\cite{koval95,gale97} 
The resulting components of the dielectric function tensor, 
which here is diagonal and also symmetric in the $ab$ plane, are reported in Table IV in 
both limits of zero frequency and high frequency. 

In Tables II and III we have also reported the corresponding values obtained by fitting to the experimental 
optical conductivity the Lorentzian model
 
\begin{equation}
\sigma_1 (\omega) = \frac{1}{60}\sum_{j=1}^{n}\frac{\omega^2 \Gamma_jS_{j}}{(\Omega^{2}_j-\omega^{2})^2+\omega^{2}\Gamma_i^2} \\
\label{sigma_fit}
\end{equation}

\noindent
In Eq. \ref{sigma_fit}, $\sigma_1 (\omega) $ is measured in $\Omega^{-1}$ cm$^{-1}$, while $\Omega_j$ and $\Gamma_j$ 
are the central frequency,   and the linewidth of the $\j$-th transverse optical mode, 
respectively, in cm$^{-1}$.  $S_j$ is  the oscillator strength in cm$^{-2}$.


\begin{table*}

\caption{\label{param} Shell-model potential parameters ($A, \rho$),
shell charges ($Y$), and on-site core-shell force constants ($k$).}  

\vskip +.2cm

\begin{ruledtabular}
\begin{tabular}{ c c c | c c c }
Interaction 
    & $A$ (eV) & $\rho$ (\AA) 
                            & Ion &  $Y$ (e) & $k$ (eV/\AA$^2$) \\
\hline
Ba-O&   915   &       0.397        
                            & Ba  &     -2    &    251       \\
Cu-O &  900   &       0.305    
                            & Cu  &      4    &      175   \\
Ge-O &  3000   &       0.280    
                            & Ge  &      0    &      1000   \\ 
     &         &          
                            & O   &      -2.59  &     50   \\                             
\label{Table I}
\end{tabular}
\end{ruledtabular}
\end{table*}

\begin{table} 
\caption{The $ab$-plane calculated phonon frequencies $\Omega_{j}$[th] and intensities  $S_{j}$[th] are compared with the frequencies $\Omega_{j}$, intensities $S_{j}$, and widths  $\Gamma_{j}$, obtained by fitting to Eq. \ref{sigma_fit} the experimental $\sigma_1(\omega)$ at 7 K  and 300 K. $\Omega_{j}$ and $\Gamma_{j}$ are in cm$^{-1}$, $S_{j}$ in cm$^{-2}$.} 
  
\begin{ruledtabular}
\begin{tabular}{cccccccccc}

Phonon ($j$) ]&  $\Omega_{j}$[th]  &  $\Omega_{j}$[7 K]& $\Omega_{j}$[300 K] & $S_{j}$[th] & $S_{j}$[7 K]  &  $S_{j}$[300 K]  &$\Gamma_{j}$[7 K]  &$\Gamma_{j}$[300 K]   \\
\colrule

1  & 59         &            &                          & 13400       &       &                  &               &                       \\
2  & 75           &          84            &    82   & 6700            &27000         &  12500       &    2     &5             \\ 
3  & 108        &         103           &  102      & 1700           &  8000         & 2300         &   2     & 3             \\
4  & 129         &                            &           & 11800          &                         &                &               &                       \\
5  & 142        &       152         &   146           & 3500        &  14500             &    9500     &    7.5    &  15   \\
6 & 179           &        187          &   182        & 2600        & 30000             & 26500   &   4      & 18          \\ 
7  & 224         &      217           &      212          & 8750   &  26000             &  19000      &    4      &  11        \\
8  & 258        &       257           &    258          & 970    & 33000          & 32000   &    21       &   25         \\
9  & 261      &       274         &      270             & 11100   & 40000           &  22600              &  6       &             15          \\
10 & 319     &      310            &   305         & 126000     &  89000          & 39000    &  11    & 20      \\
11 & 334       &       315          &  313          & 46000      &  111000        &  120000    &    4          &   15                     \\
12 & 374           &       367          &     371     & 1700        &26000             & 17000    &    8        &    30 \\
13 & 443       &                       &                  & 30                 &           &              &                &                      \\
14 & 489        &                       &                     & 450     &                 &             &                      \\
15 & 766       &      710           & 707          & 21500      &  213000         &     153000            &       7        &       17        \\
16 & 776     &     714          &   724          & 64000         &64000              & 42000    &    10      &    23        \\
17 & 786       &       772         &  772         & 2500      &  5300         &     3000   &       8      & 8      \\
18 & 836      &       844         &     838         &16500     &24000        & 12000        &     7        &     10    \\

\label{Table II}
\end{tabular}
\end{ruledtabular}
\end{table}


\begin{table} 
\caption{The $c$-axis calculated phonon frequencies $\Omega_{j}$[th] and intensities  $S_{j}$[th] are compared with the frequencies $\Omega_{j}$, intensities $S_{j}$, and widths  $\Gamma_{j}$, obtained by fitting to Eq. \ref{sigma_fit} the experimental $\sigma_1(\omega)$ at 7 K  and 300 K. $\Omega_{j}$ and $\Gamma_{j}$ are in cm$^{-1}$, $S_{j}$ in cm$^{-2}$.} 
  
\begin{ruledtabular}
\begin{tabular}{ccccccccc}
 
Phonon ($j$)  &  $\Omega_{j}$[th]   & $\Omega_{j}$[7 K]& $\Omega_{j}$[300 K] & $S_{j}$[th] & $S_{j}$[7 K] & $S_{j}$[300 K] & $\Gamma_{j}$[7 K] & $\Gamma_{j}$[300 K]      \\
\colrule

1           &     137       &    109    &   100  &      19400   &  17000     &  24000   &     7  &  22  \\ 
2           &     157       &    130    &   131  &     5600    &  6500        &   7500   &    5   &  10  \\ 					
3           &     198       &    147    &   146  &   43000    &    121000  &  66000   &     3  &  16   \\
4           &     267       &    278    &   269    &     1600  &      1900    &    3300  &     4  & 19    \\  
5           &     318       &    321    &    320    &     100      &  700       &    300     &    10    & 12   \\
6           &	     412      &    390    &    391   &     21600  & 58000      &  30000   &    15  &  19   \\
             &                 &    448    &     446 & 	            &  36000     &   43000  &   11   & 27    \\
7           &     488      &    488    &     485  &    56400  &  74000     &  47000   &  9     &  18   \\
8           &     559      &              &             &    8500     &                &               &       &         \\    
9           &     781      &    775    &    771   &      32000  &  174000    &   91000  &    7   &    9  \\
10         &     794      &    791    &    786  &     41800   &   84000    &  81000   &  17    &  21   \\
\label{Table III}
\end{tabular}
\end{ruledtabular}
\end{table}


\begin{table} 
\caption{Calculated values of the dielectric constant $\epsilon_0$ and of the high frequency dielectric function $\epsilon_{\infty}$ in the $ab$ plane and along the $c$ axis.} 
  
\begin{ruledtabular}
\begin{tabular}{cccc}
 
$\epsilon_0^{ab}$  &  $\epsilon_0^{c}$  & $\epsilon_{\infty}^{ab}$ & $\epsilon_{\infty}^{c}$  \\

\colrule

10.34     &   4.67   &   1.96        &    1.77   \\ 
\label{Table IV}
\end{tabular}
\end{ruledtabular}
\end{table}


\begin{figure}[b]
\begin{center}
{\hbox{\epsfig{figure=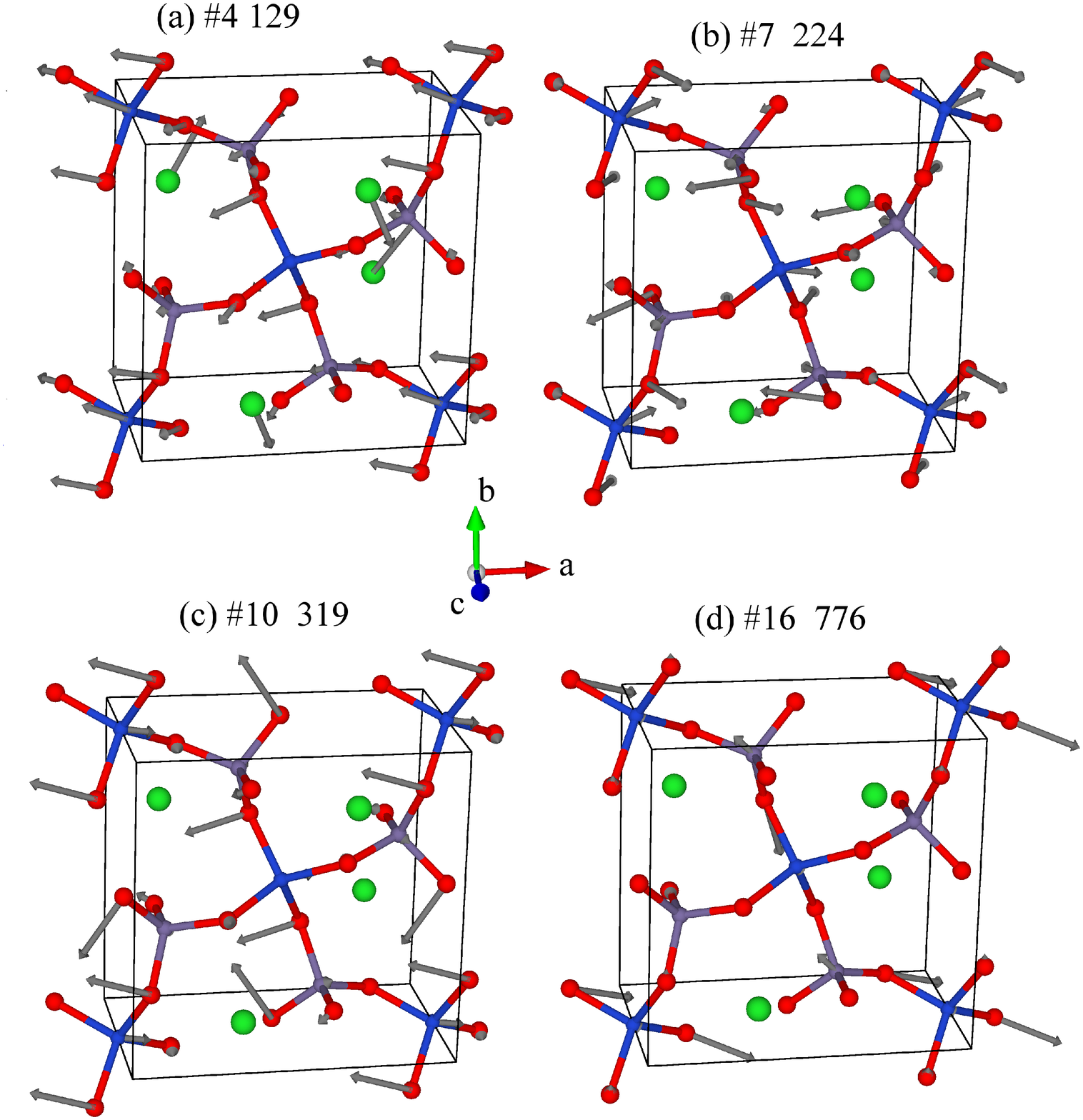,width=12cm}}}
\caption{Color online. Atomic displacements for selected phonon  modes of Ba$_2$CuGe$_2$O$_7$ along the $ab$-plane. Each mode is identified by its number in Table II and by its calculated frequency, in cm$^{-1}$.}
\label{displ_ab}
\end{center}
\end{figure}

As one can see in Tables II and III as well as in Figs. \ref{sigma_ab} and \ref{sigma_c}, 
there is a good agreement -  within the experimental linewidths - between 
the observations and the theoretical predictions, as far as the phonon frequencies are concerned. 
The main discrepancies concern phonons 15 and 16 of the $ab$ plane and modes 1-3 of the $c$ axis, 
where the calculated frequencies are much higher than the observed ones. 
This may be ascribed either to anharmonic effects which particularly affect those modes, not taken into account in the model,
or to an overestimation of the shell-shell repulsive interactions. The latter problem may be specifically  present in the Cu-O and/or Ge-O bonds in the
case of the ab-polarized phonons 15 and 16, and in the Ba-O bond in the case of the c-polarized modes 1-3.
Concerning the oscillator strengths, most calculated values reported in the same Tables and Figures match the order of magnitude of those 
measured in the real system, but are systematically lower. This may 
be mainly due to the so called ``charged phonon''
effect\cite{Rice1979} by which phonons acquire an anomalously large
spectral weight due to interatomic movement of charge associated with
their displacement pattern. Such effect is not included in the
shell model which only takes into account on-site distortions of the
atomic cloud. This results in a transfer of spectral weight to 
the far infrared region from the electronic bands in the near infrared and the visible. Similar 
"dressed phonons" were observed in several perovskites with high polarizability \cite{Homes2001,Homes2003}  
through a strong increase of the phonon intensities at low temperature, especially for the lowest-frequency modes,
and through the failure of the $f$-sum rule 

\begin{equation}
\int_{FIR}{\sigma(\omega,T)d\omega} = const.
\label{sum}
\end{equation}

\noindent
when, as in Eq. \ref{sum}, it is restricted to the far infrared (FIR). Indeed, in Ba$_2$CuGe$_2$O$_7$ both the O and Cu ions are largely polarizable as shown by the large shell charges and the small core-shell force constants (see Table I) required by the model fitting of the observed spectra. We therefore compared with each other the intensities provided by the fits at the lowest and highest temperatures of the experiment. As shown in Tables II and III, most phonons are more intense at low temperature, so that $\sum_j S_{j}$[7 K] is larger than $\sum_j S_{j}$[300 K] by about 40 $\%$ for  the $ab$ plane and 45 $\%$ for the $c$ axis. This effect, which "dresses" the phonons, is not taken into account in the shell model here employed, and may explain most of the discrepancies between theory and experiment which emerge in Tables II and III. The $f$-sum-rule violation has also the interesting implication that the dielectric constant of Ba$_2$CuGe$_2$O$_7$, which is related to $\sum_j S_{j}$ \cite{Homes2001}, is higher than  the theoretical values in Table IV and should also considerably increase when cooling the system.   As the lowest temperature here reached is larger than $T_N$, this effect would not be related to the helical spin ordering, but to a redistribution of the charges in the unit cell.

The atomic displacements for selected phonons of the $ab$ plane are shown in Fig.\ref{displ_ab}. They are labeled
by their number in Table II and by the theoretical frequency in cm$^{-1}$. Surprisingly,
we find that in many modes the CuO tetrahedra and the GeO tetrahedra have a similar pattern, 
and this is the case of the four modes shown in Fig.\ref{displ_ab}. This behavior is probably due to the similarity
of the Ge and the Cu mass which makes the tetrahedral "molecules" to vibrate at similar frequencies and to mix appreciably their modes in the
solid. Therefore, we shall discuss the pattern by referring to the tetrahedra without specifying the central atom. 

\begin{figure}[b]
\begin{center}
{\hbox{\epsfig{figure=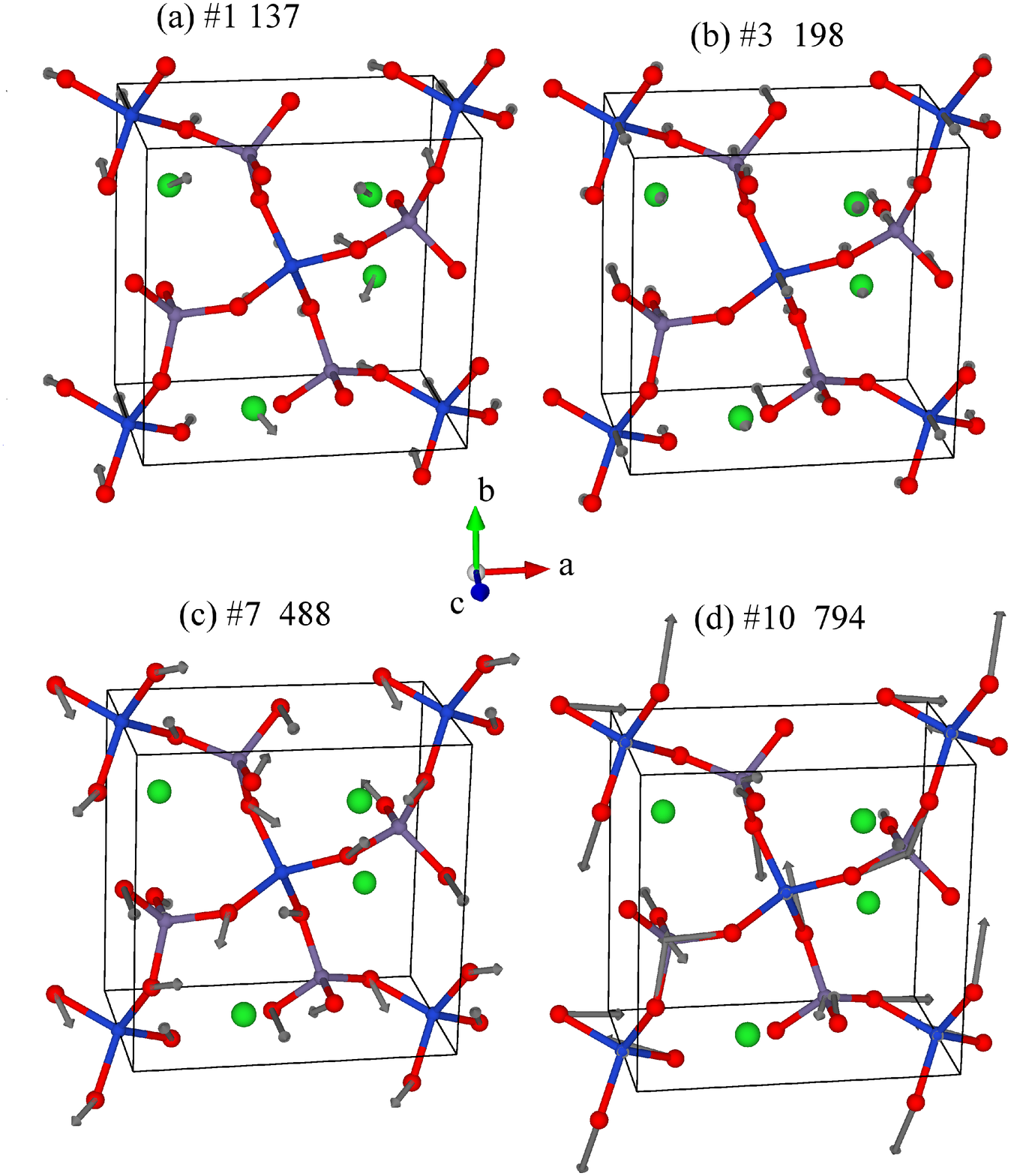,width=12cm}}}
\caption{Color online. Atomic displacements for selected  modes of Ba$_2$CuGe$_2$O$_7$ along the $c$ axis. Each mode is identified by its number in Table III and by its calculated frequency, in cm$^{-1}$.}
\label{displ_c}
\end{center}
\end{figure}

In the 129 cm$^{-1}$ mode (a) the atom at the center of the tetrahedra and three vertex oxygens move approximately in the same direction
while a fourth vertex oxygen moves in a quasi-perpendicular direction. Thus the angle of only one Cu-O (or Ge-O) bond is significantly
perturbed. In the  224 cm$^{-1}$ mode (b) two oxygens move approximately parallel to the central atom and two others move in a quasi-opposite direction, 
causing a larger distortion of the tetrahedra even if, again, mainly in the internal angles.  In the 319 cm$^{-1}$ mode
(c) the pattern is again similar for the Cu and Ge tetrahedra but they move out of phase. The central Cu moves toward a vertex O when the central Ge
moves away from a vertex O. As the other atoms move perpendicular to the bonds,  they mainly change their angles, causing the mode to be a mixture of
bending and stretching. This explains its relatively large energy. The 776 cm$^{-1}$ mode (d) has a stronger stretching
character as shown by its high energy, although one never finds in BCGO either purely stretching or purely bending modes, as it is the case in higher-symmetry solids. This also explains why here, at variance with many oxides where the highest vibrational frequency corresponds to a pure oxygen stretching mode, the highest-energy lines do not shift appreciably when decreasing the temperature.

Finally, Fig. \ref{displ_c} shows selected phonon modes polarized along the $c$ axis. They are labeled
by their number in Table III and theoretical frequency in cm$^{-1}$. Mode (a) at 137 cm$^{-1}$ shows a rigid motion of the Cu tetrahedra
with the Ba in antiphase. The Ge ions practically do not move and act as
"nodes" of the vibration. In most other modes, instead, the two kinds of tetrahedra show
very similar displacements, as discussed above for the $ab$ plane.
In mode (b) at 198 cm$^{-1}$ the Cu and Ge  tetrahedral "molecules"
move in antiphase and the Ba ions have smaller displacements. Mode (c)
at 488 cm$^{-1}$ is a mixture of stretching and bending, similar
to mode  (c) of Fig. 5. Finally, in mode (d) at 794 cm$^{-1}$, the
stretching character of the bonds bridging the O's to the center of the tetrahedra prevails.

\section{Conclusion}
In the present work we have studied the lattice dynamics of Ba$_2$CuGe$_2$O$_7$, by infrared reflectivity measurements 
with polarized radiation,
down to temperatures close to those where a helimagnetic phase takes place via the Dzyaloshinsky-Moriya mechanism. 
The number of the observed phonon lines is lower (for the $ab$ plane) 
or equal (for the $c$ axis) to that predicted for the P\=42$_1$m cell of Ba$_2$CuGe$_2$O$_7$, and no linesplitting 
has been observed when cooling the sample to 7 K. Therefore, our spectra confirm that 
the tetragonal symmetry is conserved 
down to the lowest temperatures, with no appreciable orthorhombic distortion. The optical conductivity extracted 
from $R(\omega)$ has been fit by a sum of Lorentzians, and their parameters have been compared with the results 
of  shell-model calculations. These have correctly predicted the observed frequencies, within 
the experimental linewidths, except for a few modes where the theoretical values are systematically higher. 
The discrepancy may be due either to anharmonic effects, not taken into account in the model, 
or to an overestimation of the shell-shell repulsive interactions. 
A systematic underestimation with 
respect to the observed values is instead exhibited by the calculated oscillator strengths.  
We have tentatively explained this effect by considering  that 
``charged-phonon'' effects - not considered 
in the model - can increase the dipole moment of those vibrations, due to the distortion 
of the electron clouds going beyond one atom. Such interpretation is
consistent both with the strong increase observed in the phonon intensities at low temperature, 
and with the failure of the optical sum rule when it is restricted to the phonon region. Although we cannot
make a quantitative estimate, because of lack of experimental
information in the Terahertz region and below, the remarkable increase in the phonon intensities that we observe at temperatures which are low, but higher than $T_N$,  suggests that the dielectric constant of  Ba$_2$CuGe$_2$O$_7$ may increase at those temperatures, for a redistribution of the electric charge within the cell. Measurements of the static dielectric constant could verify this expectation.

\acknowledgments

We thank Paolo Barone for suggesting software tools useful to the Figure preparation. SK thanks J. Gale and R. Migoni for useful discussions, and
acknowledges support from the Consejo Nacional de Investigaciones Cient\'{\i}ficas y T\'ecnicas de la Rep\'ublica Argentina.

\end{document}